\documentclass[11pt]{article}
\usepackage{epsf,epsfig,float,amssymb,latexsym,amsmath,amsthm,fancyhdr}
\usepackage{graphics,psfrag,longtable}
\newcommand{\vp}{\mathbf{p}}
\newcommand{\bl}{\begin{aligned}}
\newcommand{\el}{\end{aligned}}

\newcommand{\ra}{\rangle}

\newcommand{\vk}{{\mathbf{k}}}

\newcommand{\Ima}{{\rm Im}\,}
\newcommand{\Rea}{{\rm Re}\,}

\newcommand{\be}{\begin{equation}}
\newcommand{\ee}{\end{equation}}
\newcommand{\ba}{\begin{eqnarray}}
\newcommand{\ea}{\end{eqnarray}}
\newcommand{\bg}{\begin{align}}
\newcommand{\egg}{\end{align}}

\newcommand{\nn}{\nonumber}

\newcommand{\ve}{\epsilon}

\begin{document}

\title{The $\phi(1020) \, a_0(980)$ S-wave scattering and hints for a new 
 vector-isovector resonance}

\author{L.~Alvarez-Ruso$^1$, J.~A. Oller$^2$ and J.~M. Alarc\'on$^2$ \\
$^1$~Centro de F\'{\i}sica Computacional, Departamento de F\'{\i}sica, Universidade de Coimbra, Portugal\\
$^2$~Departamento de F\'{\i}sica, Universidad de Murcia, E-30071 Murcia, Spain }

\maketitle

\begin{abstract}
\noindent
We have studied the $\phi(1020)a_0(980)$ S-wave scattering at threshold energies employing chiral Lagrangians coupled to vector mesons by minimal coupling. 
The interaction is described without new free parameters by considering the scalar isovector $a_0(980)$ resonance as  dynamically generated in coupled channels, and demanding that the recently measured $e^+e^-\to\phi(1020)f_0(980)$ cross section is reproduced. For some realistic choices of the parameters, the presence of a dynamically generated isovector  companion of the $Y(2175)$ is revealed. We have also investigated the corrections to the $e^+e^-\to \phi(1020)\pi^0\eta$ reaction cross section that arise 
from $\phi(1020)a_0(980)$ re-scattering in the final state. They are  typically large and modify substantially the cross section. For a suitable choice of parameters, the presence of the resonance would manifest itself as a clear peak at $\sqrt{s}\sim 2.03$~GeV in $e^+e^-\to \phi(1020)\pi^0\eta$.   
\end{abstract}

%-----------------------------------------------------------------------------
\section{Introduction}
\label{sec:intro}
\def\theequation{\arabic{section}.\arabic{equation}}
\setcounter{equation}{0}

Our understanding of light hadron spectroscopy has been challenged in recent years by the discovery of several exotic states that cannot be easily accommodated into the quark model picture~\cite{Zhu:2007wz}. One of them is the resonance $\phi (2170)$~\cite{Amsler:2008zzb} (or $Y(2175)$, as we will refer to it from now on). The $Y(2175)$ was first observed by the BABAR Collaboration~\cite{babar1,babar2} with mass $M_Y=2175 \pm 10 \pm 15$~MeV and width $\Gamma_Y=58 \pm 16 \pm 20$~MeV~\cite{babar1} in the  $e^+e^-\to\phi(1020)\,f_0(980)$ reaction, and also found by BES in $J/\Psi \to \eta\,\phi(1020)\,f_0(980)$ decay with $M_Y=2186\pm 10 \pm 6$~MeV and $\Gamma_Y=65\pm 23\pm 17$~MeV~\cite{bes08}. The Belle Collaboration  has performed the most precise measurements up to now of the reactions $e^+e^-\to \phi(1020) \pi^+\pi^-$ and $e^+e^-\to\phi(1020)f_0(980)$ finding  $M_Y=2079\pm 13^{+79}_{-28}$~MeV and $\Gamma_Y=192\pm 23^{+25}_{-61}$~MeV~\cite{belle}. The obtained width is larger than in previous measurements but the error is large. The same feature has been found in a combined fit to both BABAR and Belle data on $e^+e^-\to \phi(1020) \pi^+\pi^-$ and $e^+e^-\to\phi(1020)f_0(980)$, yielding $M_Y=2117^{+0.59}_{-0.49}$~MeV and $\Gamma_Y=164^{+69}_{-80}$~MeV~\cite{Shen:2009mr}. 

These experimental findings have triggered a significant theoretical activity aimed at unraveling the nature and properties of this resonance. It has been interpreted as a tetraquark~\cite{wang,hosaka,polosa}, with a mass of $2.21 \pm 0.09$~GeV~\cite{wang} or $2.3\pm 0.4$~GeV~\cite{hosaka} calculated using QCD sum rules. It has also been identified with the lightest hybrid $s \bar{s}g$ state~\cite{dingyan1} with mass in the range 2.1-2.2~GeV~\cite{isgu,closest} and a width of 100-150~MeV~\cite{dingyan1}. Conventional  $s \bar{s}$ states in $2^3D_1$ or $3^3S_1$ configurations have been considered as their masses are expected to be compatible with that of the $Y(2175)$. The width of the $2^3D_1$ state is estimated  to be $150$-$210$~MeV~\cite{dingyan2} while the $3^3S_1$ is disfavored due to the rather large ($\sim 380$~MeV~\cite{blacky}) predicted width. The large width obtained in Ref.~\cite{Coito:2009na} also makes the interpretation of the  $Y(2175)$ as a dynamically generated excited state of the $\phi(1020)$ meson, within the multichannel generalization of the resonance-spectrum expansion model~\cite{beve1,beve2}, quite unlikely but further improvements of the model might change this conclusion~\cite{Coito:2009na}. Reference~\cite{torres} studies the three-body $K\bar{K}\phi(1020)$ scattering with two-body interactions taken from unitarized chiral perturbation theory \cite{npa,roca} and  a resonance with 2170~MeV of  mass but a width of  only 20~MeV  is generated. 
In Ref.~\cite{fif0} we obtained a good description of the $e^+e^-\to \phi(1020) f_0(980)$ scattering data in the threshold region ($\sim 2$~GeV) using chiral Lagrangians coupled to vector mesons, supporting the conclusion that the $Y(2175)$ has a large mesonic $\phi(1020) f_0(980)$ component.    

In this contest, it is relevant to establish whether there is an isovector companion of the isoscalar $Y(2175)$. Such an investigation will help constraining theoretical models and their parameters, leading to a better understanding of meson properties in the energy region around 2~GeV. In particular, the Faddeev-type calculation of Ref.~\cite{torres} that obtains the $Y(2175)$ as a dynamically generated state finds no resonance in the isovector $\phi(1020) a_0(980)$ S-wave channel. Experimentally, this isovector resonance could show up in  $e^+e^-\to \phi(1020) a_0(980) \to \phi(1020) \pi^0\eta$, as suggested in a recent theoretical study of this process~\cite{araujo}. It could also be observed in the $e^+e^- \to \phi(1020) K^+ K^-$ reaction because the $a_0(980)$ couples strongly to $K^+ K^-$~\cite{newone}. One should stress that the calculations of Refs.~\cite{araujo,newone} do not take into account $\phi(1020)a_0(980)$ final state interactions (FSI) which, resonant or not, could be large and have a sizable impact on the predicted cross sections.  

In this article we apply the formalism of Ref.~\cite{fif0} to the S-wave  $\phi(1020) a_0(980)$ scattering and discuss the possible presence of an isovector $J^{PC}=1^{--}$  dynamically generated resonance around the $\phi(1020)a_0(980)$ threshold for parameters that satisfactorily describe the isoscalar $\phi(1020) f_0(980)$ channel. FSI corrections to $e^+e^-\to \phi(1020) a_0(980) \to \phi(1020) \pi^0\eta$ are also studied. The formalism for $\phi(1020)a_0(980)$ scattering is developed in Sec.~\ref{sec:ff} followed by the derivation of the scattering amplitudes. Section~\ref{sec:fb} contains the results and discussions thereof. Our concluding remarks are given in Sec.~\ref{sec:conclu}.

%------------------------------------------------------------------------------
\section{Derivation of the $\phi(1020) a_0(980)$ scattering amplitude}
\label{sec:ff}
\def\theequation{\arabic{section}.\arabic{equation}}
\setcounter{equation}{0}

In order to obtain the  $\phi(1020) a_0(980)$ amplitude we follow closely our previous paper \cite{fif0}, replacing the isoscalar $f_0(980)$ by the isovector $a_0(980)$. First, the scattering of the  $\phi(1020)$ resonance with an S-wave neutral pair of the lightest pseudoscalar mesons in isospin $I=1$ is investigated. Two types of  meson pairs are then possible, namely, $|1\ra \equiv |K\bar{K}\ra_{I=1}$ and $|2 \ra \equiv |\pi^0 \eta_8\ra$ (already a pure $I=1$ state.) The following channels result
\begin{align}
&(1 \to 1): \qquad \phi(1020)\, |K\bar{K}\ra_{I=1} \to \phi(1020)\, |K\bar{K}\ra_{I=1} \,,\nn \\
&(1 \to 2): \qquad \phi(1020)\, |K\bar{K}\ra_{I=1} \to \phi(1020)\, |\pi^0 \eta_8\ra \,,\nn \\
&(2 \to 1): \qquad \phi(1020)\, |\pi^0 \eta_8\ra \to \phi(1020)\, |K\bar{K}\ra_{I=1} \,,\nn \\
&(2 \to 2): \qquad \phi(1020)\, |\pi^0 \eta_8\ra \to \phi(1020)\, |\pi^0 \eta_8\ra \,.
\label{channels}
\end{align}

\begin{figure}[h]
\psfrag{k}{$k$}
\psfrag{p}{$p$}
\psfrag{l}{$\ell$}
\psfrag{pi}{$\pi$}
\psfrag{r}{$k-\ell$}
\centerline{\epsfig{file=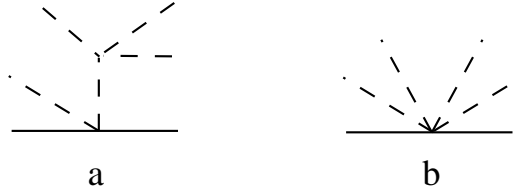,width=.4\textwidth,angle=0}}
\vspace{0.2cm}
\caption[pilf]{\protect \small
Dominant tree-level contributions to the scattering of a $\phi(1020)$ with a neutral pair of pseudoscalar mesons close to threshold. The dashed lines denote the pseudoscalar mesons and the solid line stands for the $\phi(1020)$.
\label{fig:pole0}}
\end{figure}

Both $\phi(1020)a_0(980)$ S- and D- waves contribute to the $1^{--}$ channel but since we are interested in the threshold region around  2~GeV, D-wave terms can be neglected. They are suppressed by powers of  $|\vp|^{2n}$, where  $|\vp|$ is the three momentum of the $\phi(1020) a_0(980)$ pair in the center of mass (CM),  and $n=1,2$ is the number of  $\phi(1020)a_0(980)$ D-wave  initial and final states involved. Moreover, as both $\phi(1020)$ and $a_0(980)$ are very close to the $K\bar{K}$ threshold,  the amplitude at tree level is dominated by diagram Fig.~\ref{fig:pole0}a. The main reason is that the propagator of the kaon intermediate state is almost on-shell.\footnote{In our case, the intermediate state is a kaon or an anti-kaon because of the absence of $\phi^2\pi^0\pi^0$, $\phi^2\pi^0\eta_8$ and $\phi^2\eta_8\eta_8$ vertices in the $V^2\Phi^2$ Lagrangian of Eq.~(\ref{vvff}). See Ref.~\cite{fif0} for a detailed analysis where all the other tree-level diagrams originating from the same set of Lagrangians employed are discussed and shown to be suppressed compared to Fig.~\ref{fig:pole0}a.} We also include the local  term of Fig.~\ref{fig:pole0}b because  the off-shell part of the four-pseudoscalar-meson vertex  can cancel the kaon propagator generating 
local terms. Therefore one has to consider simultaneously the sum of amplitudes from both diagrams as any splitting would depend on field parameterization. 

The required vertices 
  can be obtained from the lowest order SU(3) chiral Lagrangian~\cite{glsu3}
\begin{align}
{\cal L}_2=\frac{f^2}{4}\hbox{Tr}\left(D_\mu U^\dagger D^\mu U+ \chi^\dagger U+\chi U^\dagger \right)\,,
\label{lag2}
\end{align}
with $f$ the pion weak decay constant in the chiral limit, that we approximate to $f_\pi=92.4$~MeV. 
The octet of the lightest pseudoscalar fields are included in $U$ as 
\begin{align}
U&=\exp\left(i\frac{\sqrt{2}\Phi}{f}\right) \,,\nn\\
\Phi&=\frac{1}{\sqrt{2}}\sum_{i=1}^8 \phi_i\lambda_i=\left(
\begin{array}{ccc}\frac{\pi^0}{\sqrt{2}}+\frac{1}{\sqrt{6}}\eta_8 & \pi^+ & K^+\\ 
\pi^- & -\frac{\pi^0}{\sqrt{2}}+\frac{1}{\sqrt{6}}\eta_8 & K^0\\
K^- & \overline{K}^0 & -\frac{2}{\sqrt{6}}\eta_8
 \end{array}\right)\,.
 \label{uu3}
 \end{align}
Assuming minimal coupling, the covariant derivative $D_\mu U$ incorporates the lightest octet of vector resonances $v_\mu$ as external fields: 
\begin{align}
D_\mu U&=\partial_\mu U-i g \left[ v_\mu, U \right] \,, \nn\\
v_\mu&=\left(
\begin{array}{ccc}
\frac{\rho^0}{\sqrt{2}}+\frac{1}{\sqrt{2}}\omega& \rho^+ & {K^*}^+\\
\rho^- & -\frac{\rho^0}{\sqrt{2}}+\frac{1}{\sqrt{2}}\omega & {K^*}^0\\
{K^*}^- & {{\overline{K}^*} }^0 & \phi
\end{array}
\right) \,,
\label{matv}
\end{align}
where $g$ is a coupling constant.  We have assumed ideal mixing, so that $\phi=-\sqrt{\frac{2}{3}}\omega_8+\frac{1}{\sqrt{3}}\omega_1$
 and $\omega=\frac{1}{\sqrt{3}}\omega_8+\sqrt{\frac{2}{3}}\omega_1$, with $\omega_8$ and $\omega_1$ being the $I=0$ octet and singlet vector states.
As a result, the following Lagrangians involving vector and pseudoscalar mesons arise from Eq.~\eqref{lag2}: 
\begin{align}
{\cal L}_{V^2\Phi^2}&=g^2\hbox{Tr}\left( v_\mu v^\mu \Phi^2-v_\mu \Phi v^\mu \Phi\right)\,,\nn\\
{\cal L}_{V^2\Phi^4}&=-\frac{g^2}{6f^2}\hbox{Tr}\left( v_\mu v^\mu \Phi^4-4 v_\mu
\Phi^3v^\mu \Phi+3 v_\mu \Phi^2 v^\mu \Phi^2\right)\,, \nn\\
{\cal L}_{\Phi^4}&=-\frac{1}{6f^2}\hbox{Tr}\left(\partial_\mu \Phi \partial^\mu \Phi \Phi^2 - \partial_\mu \Phi \Phi \partial^\mu \Phi \Phi -\frac{1}{2} M \Phi^4 \right)\,, 
\label{vvff}
\end{align}
where $M=\mathrm{diag}(m_\pi^2,m_\pi^2,2m_K^2-m_\pi^2)$ and $m_\pi$ and $m_K$ are the pion and kaon masses.  
To construct the $(1 \to 1)$ amplitude we cast $|K\bar{K}\ra_{I=1}$ as
 \begin{align} 
|K(k_1)\bar{K}(k_2)\ra_{I=1}&=-\frac{1}{\sqrt{2}}|K^+(k_1)K^-(k_2)-K^0(k_1)\bar{K}^0(k_2)\ra\,,
\label{isostate}
 \end{align}
 with $k_1$ and $k_2$ the kaon four-momenta.  
The global minus sign appears because we identify $|K^-\ra=-|I=1/2,I_3=-1/2\ra$ to be consistent with the convention adopted in the chiral Lagrangians Eq.~(\ref{uu3}). We denote the amplitudes for the reaction channels
 \begin{align}
\phi(p) \, K^+(k_1) \, K^-(k_2) &\to \phi(p') \,K^+(k'_1) \,K^-(k'_2) \,,\nn\\
\phi(p) \, K^0(k_1) \, \bar{K}^0(k_2) &\to \phi(p') \,K^+(k'_1) \,K^-(k'_2) \,,\nn\\
\phi(p) \,K^+(k_1) \,K^-(k_2) &\to \phi(p') \,K^0(k'_1) \,\bar{K}^0(k'_2) \,,\nn\\
\phi(p) \,K^0(k_1) \,\bar{K}^0(k_2) &\to \phi(p') \,K^0(k'_1) \, \bar{K}^0(k'_2) \,,
\end{align}
as $T_{cc}$, $T_{nc}$, $T_{cn}$ and $T_{nn}$, from top to bottom. These amplitudes were calculated in Ref.~\cite{fif0} for the diagrams of Fig.~\ref{fig:pole0} assuming isospin symmetry. The result for diagram Fig.~\ref{fig:pole0}a is
\begin{align}
T^{(a)}_{cc}&=-\frac{2 g^2}{3f^2} \ve\cdot \ve' \left\{4 + 6 (u_a-2m_K^2)\Bigl[D(Q+k_1)+D(Q-k'_2)\Bigr] \right. \nn \\
&\left. + 6 (u_b-2m_K^2)\Bigl[D(Q+k_2)+D(Q-k'_1)\Bigr] \right\}  \,,\nn\\
T_{nn}^{(a)}&=T^{(a)}_{cc} \,,\nn \\
T_{nc}^{(a)}&=\frac{1}{2}T^{(a)}_{cc} \,, \nn \\
T_{cn}^{(a)}&=T_{nc}^{(a)} = \frac{1}{2}T^{(a)}_{cc} \,,
\label{tcn}
\end{align}
where $\epsilon$ ($\epsilon'$) is the polarization four-vector of the initial (final) $\phi(1020)$ meson, $u_a=(k'_1-k_2)^2$, $u_b=(k'_2-k_1)^2$ and $Q=p-p'$.   The kaon propagator $D(q)$ is given by
\begin{align}
D(q)=\frac{1}{m_K^2-q^2-i\varepsilon}\,, 
\end{align}
with $\varepsilon\to 0^+$. For the contact term (diagram  of Fig.~\ref{fig:pole0}b) the result is
\begin{align}
T_{cc}^{(b)}&=-\frac{16g^2}{3f^2}\ve\cdot \ve' \,,\nn\\
T_{nn}^{(b)}&=T_{cc}^{(b)} \,,\nn\\
T^{(b)}_{nc}&=T^{(b)}_{cn}=\frac{1}{2}T_{cc}^{(b)} \,.
\end{align}
Taking into account  Eq.~\eqref{isostate} one finds for the $(1 \to 1)$ channel of Eq.~(\ref{channels})  
\begin{align}
T^{I=1}_{11}&=\frac{1}{2}\left\{T_{cc}+T_{nn}-T_{cn}-T_{nc}\right\}=\frac{1}{2}T_{cc} \,,
\label{t2i0}
\end{align}
where $T_{cc}=T_{cc}^{(a)}+T_{cc}^{(b)}$ (and analogously for $T_{nn}$, $T_{cn}$ and $T_{nc}$). Therefore, 
\begin{align}
T^{I=1}_{11}=\frac{2 g^2}{f^2}\ve\cdot \ve' \Bigl\{ -2 + k_2\cdot k'_1\bigl[D(Q+k_1)+D(Q-k'_2)\bigr]+k_1\cdot k'_2 \bigl[D(Q+k_2)+D(Q-k'_1)\bigr] \Bigr\} \,.
\label{ti1}
\end{align}

Proceeding in the same way, the $(1 \to 2)$ and $(2 \to 1)$ amplitudes are found to be
\begin{align}
%T^{I=1}_{12} & =-\frac{2 g^2}{\sqrt{6} f^2}\ve\cdot \ve' \left[ 3 k'_1\cdot %k'_2 +\frac{4}{3} (m_{\pi}^2 -m_K^2) + m_\eta^2 \right]
%\left[ D(Q+k_1)+D(Q+k_2) \right]\,, \nn \\
%T^{I=1}_{21} & =-\frac{2 g^2}{\sqrt{6} f^2}\ve\cdot \ve' \left[ 3 k_1\cdot k_2 %+\frac{4}{3} (m_{\pi}^2 -m_K^2) + m_\eta^2 \right]
%\left[ D(Q-k'_1)+D(Q-k'_2) \right]\,,
T^{I=1}_{12} & =-\frac{2 g^2}{\sqrt{6} f^2}\ve\cdot \ve' \left[ 3 k'_1\cdot k'_2  + m_\pi^2 \right]
\left[ D(Q+k_1)+D(Q+k_2) \right]\,, \nn \\
T^{I=1}_{21} & =-\frac{2 g^2}{\sqrt{6} f^2}\ve\cdot \ve' \left[ 3 k_1\cdot k_2 + m_\pi^2 \right]
\left[ D(Q-k'_1)+D(Q-k'_2) \right]\,,
\label{extra.t}
\end{align}
where $k_1$, $k_2$ ($k'_1$, $k'_2$) are the four-momenta of the initial (final) pseudoscalars. 
The Gell-Mann-Okubo mass relation $m_\eta^2=4 m_K^2/3-m_\pi^2/3$ has been used to simplify the final expressions. Notice that there is no local term due to a cancellation between the contact term from Fig.~\ref{fig:pole0}b and the local part from Fig.~\ref{fig:pole0}a. Finally, $(2 \to 2)$ is absent at tree level because there are no $\phi^2 \Phi^2$ or  $\phi^2 \Phi^4$ vertexes with only $\pi^0$ and $\eta$ mesons. Because of the absence of the contact terms for $(1\to 2)$, $(2\to 1)$ and $(2\to 2)$ there is no need to further consider these processes in order to obtain the $\phi(1020)a_0(980)$ interaction kernel. It can be obtained directly from $(1\to 1)$, as it is explicitly worked out below.

\begin{figure}[h]
\psfrag{f0}{{\tiny $a_0(980)$}}
\centerline{\epsfig{file=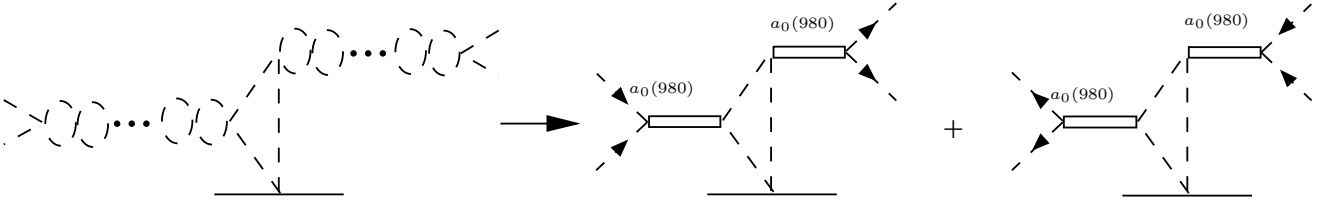,width=1.\textwidth,angle=0}}
\vspace{0.2cm}
\caption[pilf]{\protect \small
The two $a_0(980)$ poles that arise from meson-meson interactions in $I=1$ and S-wave. 
 \label{fig:itera}}
\end{figure}
Next, we consider initial- and final-state re-scattering of the pseudoscalar mesons in $I=1$ and S-wave from the diagrams in Fig.~\ref{fig:pole0}, as shown in Fig.~\ref{fig:itera} for the nonlocal part of the interaction. The re-scattering chains, made of $K\bar{K}$ and $\pi^0 \eta$ pairs, contain the poles of the initial and final $a_0(980)$\ resonances~\cite{Weinstein:1990gu,Janssen:1994wn,npa,nd,mixing}.
 Below, the residue at the $a_0(980)$ double pole will be identified as the $a_0(980)\phi(1020)$ interaction kernel ${\mathcal K}_{\phi a_0}$. 
  We follow Refs.\cite{npa,nd,mixing}, where the $I=1$ S-wave meson-meson scattering was studied with $K\bar{K}$ and $\pi^0\eta$ coupled channels, and the $a_0(980)$ resonance was dynamically generated from the meson-meson self-interactions. This conclusion is also shared 
with  other approaches like Refs.\cite{Janssen:1994wn,Weinstein:1990gu}.
The $I=1$ S-wave meson-meson amplitudes $T_{ij}$ fulfill the Bethe-Salpeter equation in coupled channels \cite{npa,nd}
\begin{equation}
T_{ij} =  \sum_{m} {\mathcal K}_{im} \left( \delta_{mj} - G_m T_{mj} \right)\,,
\label{BS} 
\end{equation}
where  the indices $i,j,m=1,~2$ denote the $K\bar{K}$ and $\pi^0 \eta$ channels. The $T$-matrix is given in terms of the on-shell part of the $I=1$ S-wave meson-meson amplitudes at tree level ${\cal K}_{ij}$ and the $K\bar{K}$ and $\pi^0 \eta$ unitary scalar loop functions, $G_1$ and $G_2$ in this order.\footnote{The expressions for $G_{\phi a0}$ given in Eqs.~(\ref{gff}) and (\ref{g.cut}), obtained from a dispersion relation and with cut-off regularization, respectively, are also applicable here after the appropriate replacement of masses.} Notice that the ${\cal K}_{ij}$ factorize  in Eq.~\eqref{BS}  \cite{npa}. They are calculated from ${\cal L}_{\Phi^4}$, Eq.~(\ref{vvff}), with the resulting expressions~\cite{npa}
\begin{align}
{\cal K}_{11}&\equiv {\cal K}_{K \bar{K}\to K\bar{K}}=\frac{k^2}{4f^2} \,, \nn \\
{\cal K}_{12}={\cal K}_{21}&\equiv {\cal K}_{\pi^0\eta\to K\bar{K}}=-\frac{\sqrt{3/2}}{6 f^2}(3 k^2-4m_K^2)\,,\nn\\
{\cal K}_{22}&\equiv {\cal K}_{\pi^0\eta\to \pi^0\eta}=\frac{m_\pi^2}{3f^2}\,,
\label{v22.eq}
\end{align}
with $k^2$ being the invariant mass squared of the meson pair.  

\begin{figure}[h]
\psfrag{l}{{\tiny $\ell$}}
\psfrag{l+k}{{ \tiny $\ell+k$}}
\psfrag{l+k'}{{\tiny $\ell+k'$}}
\psfrag{l-k1}{{\tiny $\ell-k$}}
\psfrag{l-k2}{{\tiny $\ell-k'$}}
\psfrag{Kp}{{\tiny $K^+$}}
\psfrag{(K0)}{{\tiny $(K^0)$}}
\psfrag{fi}{{\tiny $\phi$}}
\psfrag{K}{{\tiny $K$}}
\psfrag{Kb}{{\tiny $\bar{K}$}}
\centerline{\epsfig{file=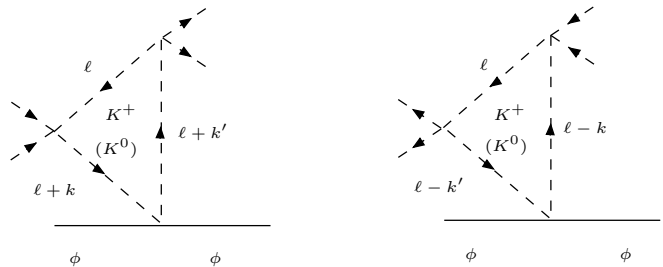,width=.5\textwidth,angle=0}}
\vspace{0.2cm}
\caption[pilf]{\protect \small
Triangular kaon-loop graphs with a $K^+$ or a $K^0$ running in the loop.  \label{fig:triangle}}
\end{figure}

In presence of re-scattering of the initial and final two-body hadronic states, the dressed amplitudes
  ${\cal M}_{ij}$ can be cast as 
\begin{equation}
{\mathcal M}_{ij} = \sum_{mn} (\delta_{im} - T_{im} G_m ) V_{mn}  (\delta_{nj} - G_n T_{nj}) \,,  
\label{IFSI} 
\end{equation}
The first (last) term in parentheses accounts for the initial (final) state interactions between the pair of pseudoscalar mesons in I=1 and S-wave. For its derivation and other applications 
see Refs.~\cite{gamma,ddecays}. The $V_{mn}$ part, which contains the $\phi(1020)$ interaction with the pseudoscalar pair projected into S-wave, consists of two terms, $V_{mn}=V^{(c)}_{mn}+V^{(t)}_{mn}$. The first one is a local term, present only in the $(1 \to 1)$ channel as shown above. From Eq.~(\ref{ti1})    
\begin{equation}
V^{(c)}_{mn}=\frac{4 g^2}{f^2} \delta_{m1} \delta_{n1} \,,
\label{local}
\end{equation}
where only the leading non relativistic contribution to $\ve(\vp,s)\cdot \ve'(\vp',s')\approx -\delta_{ss'}$ has been kept; this approximation is justified for small $\phi(1020)$ (and $a_0(980)$) velocities in the $\phi(1020)a_0(980)$ CM frame. The second term $V^{(t)}_{mn}$ is given by the triangular loop diagrams depicted in Fig.~\ref{fig:triangle} with only kaons in the internal lines. For the $\Phi^4$ vertices we take only the on-shell amplitudes of Eq.~(\ref{v22.eq}). The off-shell parts  are proportional to the inverse of kaon propagators and cancel with them in the calculation of the loop,  giving rise to amplitudes that do not correspond anymore to the dominant triangular kaon-loop but to other topologies~\cite{npa,fif0}. Nonetheless one should bare in mind that some of these sub-leading contributions may alter the contact term, fixed above from the tree level amplitudes.    
We obtain 
\begin{align}
V^{(t)}_{mn}=-4 g^2 {\mathcal K}_{m1}(k^2) {\mathcal K}_{n1}(k'^2) L_S \,,
\end{align}
where
\begin{align}
L_S =
\frac{1}{8\pi^2}\int_{-1}^{+1}\frac{d\cos\rho}{Q^2} \int_0^{1/2} dx
\frac{1}{c} \left[ \log\left(1-2x/c\right)-\log\left(1+2x/c\right) \right] \,,
\label{ms}
\end{align}
with 
\begin{align}
c^2= \frac{4}{Q^2} \left[ x^2 Q^2 + 2 k^2 x (1-2 x) - m_K^2  + i \epsilon \right] \,;
\label{c}
\end{align}
$k^2$ ($k'^2$) stand for the invariant mass squared of the initial (final) pseudoscalar-meson pair. Inside the integral we take $k^2=k'^2$, which holds at the $a_0(980)$ double pole.  We account for the S-wave projection by averaging over $\cos\rho$ with 
$\rho$ the relative angle between $\phi(1020)$ incoming ($\vp$) and outgoing ($\vp'$) momenta in the CM frame. In terms of this angle $Q^2=-2\vp^2(1-\cos\rho)$. As for the contact term we approximate $\ve(\vp,s)\cdot \ve'(\vp',s')\approx -\delta_{ss'}$. 
Further details on the derivation of the triangular-loop amplitude can be found in Ref.~\cite{fif0}.\footnote{The expression for $c$ given in Eq.~(\ref{c}), although more compact, coincides with Eq.~(2.24) of Ref.~\cite{fif0}.}  
Altogether,
\begin{equation}
V_{mn} =4 g^2 \left( \frac{\delta_{m1} \delta_{1n}}{f^2} - {\mathcal K}_{m1}(k^2) {\mathcal K}_{n1}(k'^2) L_S \right) \,.
\end{equation}

Substituting the previous expression in Eq.~(\ref{IFSI}) and using Eq.~(\ref{BS}) one finds that 
\begin{equation}
{\mathcal M}_{ij} = 4 g^2 \left\{ \frac{1}{f^2} [\delta_{i1} - T_{i1}(k^2) G_1(k^2)] [\delta_{1j} - G_1(k'^2) T_{j1}(k'^2)]
 - T_{i1}(k^2) T_{j1}(k'^2) L_S \right\} \,.
\label{Eq.ms}
\end{equation} 

Now we proceed to extract the $\phi(1020)a_0(980)$ interaction kernel. For this purpose we notice that the scattering amplitude $T_{11}(k^2)$ contains the $a_0(980)$ resonance pole with residue 
\begin{align}
\lim_{k^2\to M_{a_0}^2} (M_{a_0}^2-k^2)T_{11}(k^2)= \gamma_{K\bar{K}}^2\,.
\label{resi}
\end{align}
where $M_{a_0}^2$ denotes the $a_0(980)$ pole position. Therefore,
\begin{align}
{\cal K}_{\phi a_0}&=\frac{1}{\gamma_{K\bar{K}}^2}\lim_{k^2, k'^2\to M_{a_0}^2}(k^2-M_{a_0}^2)(k'^2-M_{a_0}^2){\cal M}_{11} \nn\\
&= 4 g^2 \gamma_{K\bar{K}}^2 
\left[\frac{1}{f^2} G_{1}(M_{a_0}^2)^2- L_S \right]\,.
\label{vff}
\end{align} 
The $1/\gamma_{K\bar{K}}^2$ factor appears because ${\cal M}_{11}$ contains two extra  $a_0(980)\to |K\bar{K}\ra_{I=1}$ couplings that should be removed in order to isolate the $a_0(980)$ resonances.    

\begin{figure}[ht]
\psfrag{GeV}{{\small $\sqrt{s}$~[GeV]}}
\centerline{\epsfig{file=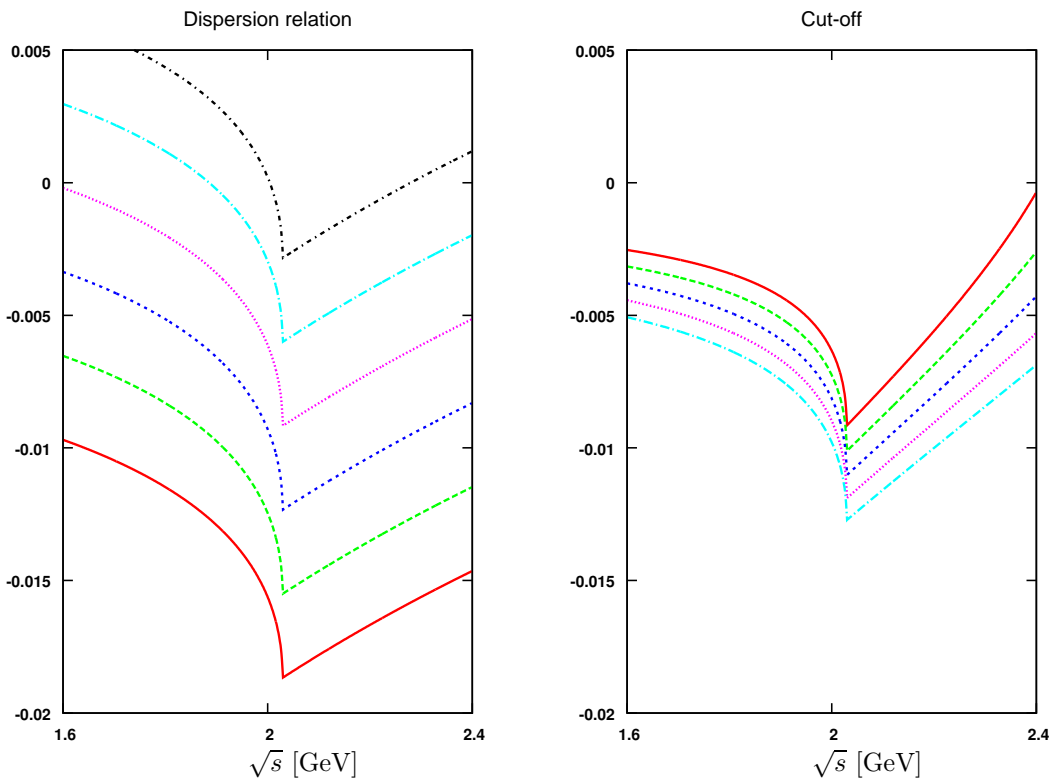,width=.6\textwidth,angle=-90}}
\vspace{0.2cm}
\caption[pilf]{\protect \small (Color online). The  real part of $G_{\phi a_0}$ calculated from a dispersion relation, Eq.~\eqref{gff}, left panel, and with a three-momentum cut-off, Eq.~\eqref{g.cut}, right panel. From top to bottom, the subtraction constant $a_1$ is varied from $-1.0$ to $-3.5$ in steps of $-0.5$  while the three-momentum cut-off, $q_{max}$, goes from 0.8 to 1.2~GeV in steps of 0.1~GeV. 
\label{fig:gs}}
\end{figure}

Finally, the  $\phi(1020)a_0(980)$ S-wave scattering amplitude is 
\begin{align}
T_{\phi a_0}=\frac{{\cal K}_{\phi a_0}}{1+{\cal K}_{\phi a_0} G_{a_0 f_0}}\,.
\label{tff}
\end{align}
For a general derivation of this equation, analogous to Eq.~\eqref{BS}, based on the N/D method see Refs.~\cite{nd,kn}. Using dispersion relations, the $\phi(1020)a_0(980)$ loop function, $G_{\phi a_0}$,  is found to be~\cite{nd}
\begin{align}
G_{\phi a_0}(s)&=\frac{1}{(4\pi)^2}\biggl\{
a_1+\log\frac{M_{a_0}^2}{\mu^2}-\frac{M_\phi^2-M_{a_0}^2+s}{2s}\log\frac{M_{a_0}^2}{M_\phi^2}
+\frac{|\vp|}{\sqrt{s}}\biggl[\log(s-\Delta+2\sqrt{s}|\vp|)\nn\\
&+\log(s+\Delta+2\sqrt{s}|\vp|)
-\log(-s+\Delta+2\sqrt{s}|\vp|)-\log(-s-\Delta+2\sqrt{s}|\vp|)
\biggr]\biggr\}\,,
\label{gff}
\end{align}
with $\Delta=M_{\phi}^2-M_{a_0}^2$. While the renormalization scale $\mu$ is fixed the to value of the $\rho$ meson mass, $\mu=770$~MeV, the subtraction constant $a_1$ has to be fitted to data~\cite{nd}.
The loop-function can also be regularized with a three-momentum cut-off $q_{max}$~\cite{npa},
\begin{align}
G_{\phi a_0}(s)=\int_0^{q_{max}}\frac{|\vk|^2 d|\vk|}{(2\pi)^2}\frac{w_{\phi}+w_{a_0}}{w_{\phi} w_{a_0}(s-(w_{\phi}+w_{a_0})^2+ i \epsilon)}\,,
\label{g.cut}
\end{align}
with  $w_i=\sqrt{m_i^2+|\vk|^2}$.\footnote{Of course, this regularization procedure spoils the  analytical properties of $G_{\phi a_0}$.}
It is instructive to compare the real part of the $G_{\phi a_0}$ functions that result from the two methods. For this we fix $M_{a_0}=1.009$~GeV, corresponding to the pole mass obtained in Ref.~\cite{npa}. The comparison is presented in Fig.~\ref{fig:gs}. On the left panel, Eq.~\eqref{gff} is evaluated varying the subtraction constant $a_1$ from $-1.0$ to $-3.5$ in steps of $-0.5$ starting from the top  while on the right one, Eq.~\eqref{g.cut} is plotted for  $q_{max}$  between 0.8 and 1.2~GeV (around the typical hadronic scale $\sim 4 \pi f_\pi$) in steps of 0.1~GeV from top to bottom. We observe a significant overlap between both functions  in the threshold region ($\sim 2$~GeV) for values of $a_1$ between $-3$ and $-2$.  This interval contains indeed the $a_1$ values obtained in Ref.~\cite{fif0} by fitting the  $e^+e^-\to \phi(1020) f_0(980)$ cross section. This coincidence is interpreted as an indication that the $Y(2175)$ resonance is to a large extent dynamically generated. Now, we  investigate this possibility for the $I=1$ S-wave  $\phi(1020)a_0(980)$ scattering. 

%%%%%%%%%%%%%%%%%%%%%%%%%%%%%%%%%%%%%%%%%%%%%%%%%%%%%
%
%%%%%%%%%%%%%%%%%%%%%%%%%%%%%%%%%%%%%%%%%%%%%%%%%%%%%%
\section{Results and discussion}
\label{sec:fb}
\def\theequation{\arabic{section}.\arabic{equation}}
\setcounter{equation}{0}

\subsection{Possible $\phi(1020)\, a_0(980)$ resonances}

In this investigation, we consider two possibilities for the $a_0(980)$ properties (pole position and residue), as they depend on the adopted approach. In the first one, the Bethe-Salpeter (BS) equation for meson-meson scattering was solved using cut-off regularization for the loop function~\cite{npa}. In the second case, the N/D method was used with the meson-meson loop function obtained with a dispersion relation~\cite{nd}. It additionally includes the s-channel exchanges of tree-level scalar resonances, corresponding to a flavor singlet of mass close to 1~GeV and a higher octet of mass around 1.4~GeV.\footnote{The $a_0(980)$ pole position obtained with the N/D approach is almost identical to the one obtained with the Inverse Amplitude method~\cite{iam}.} In both studies the $K\bar{K}$ and $\pi^0 \eta$ coupled channels were considered for $I=1$. The $a_0(980)$ properties extracted in these references are listed in Table~\ref{a0proper}.
\begin{table}[h]
\begin{center}
\begin{tabular}{|c|c|c|}
\hline
& $M_{a_0}$~[GeV] & $\gamma^2_{K \bar{K}}$~[GeV$^2$] \\
\hline
BS~\cite{npa} & $1.009+i\,0.056$  & $24.73-i\,10.82$ \\
N/D~\cite{nd} & $1.055+i\,0.025$ & $17.37-i\,24.77$ \\
\hline 
\end{tabular}
\caption{$a_0(980)$ properties, pole position $M_{a_0}$ and residue $\gamma^2_{K \bar{K}}$, used as input. 
\label{a0proper}}
\end{center}
\end{table} 
 
Furthermore, we employ two sets of values for the coupling $g$ and the $\phi a_0$ subtraction constant $a_1$  corresponding to the values we obtained in Ref.~\cite{fif0} by fitting BABAR~\cite{babar2} and Belle~\cite{belle} data on $e^+e^-\to\phi(1020)\,f_0(980)$. The first of the fits corresponds to Fit 1 of Ref.~\cite{fif0}, with mass and couplings for the $f_0(980)$ resonance 
from Ref.~\cite{alba},  while the second one is similar to Fit 2 of Ref.~\cite{fif0} but obtained with slightly different values of the $f_0(980)$ mass and $K \bar{K}$ residue ($\gamma^2_{f_0 K\bar{K}}$), corresponding to those values of Ref.~\cite{nd}. The $f_0(980)$ properties 
 from Refs.~\cite{alba,nd} and the resulting fit parameters are collected in Table~\ref{f0proper}.\footnote{The difference in the subtraction constant $a_1$ from both sets is too small to be significant.} 
Notice that   $g^2<0$. As remarked in Ref.~\cite{fif0}, $g^2$ should be understood as a parameter characterizing the $\phi(1020)K$ scattering around its threshold, with presumably large influence from the $I(J^P)=\frac{1}{2}(1^+)$ $K_1(1400)$ resonance~\cite{Amsler:2008zzb}, which would determine the negative 
sign for $g^2$. 
\begin{table}[h]
\begin{center}
\begin{tabular}{|c|c|c|c|c|}
\hline
& $M_{f_0}$~[GeV] (fixed) & $\gamma^2_{f_0 K\bar{K}}$~[GeV$^2$] (fixed) & $\sqrt{-g^2}$ & $a_1$ \\
\hline
Fit 1 & 0.980 & 16 & $7.33 \pm 0.30$ & $-2.41 \pm 0.14$ \\
Fit 2 & 0.988 & 13.2 & $5.21 \pm 0.12$ & $-2.61  \pm 0.14$ \\
\hline 
\end{tabular}
\caption{Fits to  BABAR~\cite{babar2} and Belle~\cite{belle}   $e^+e^-\to\phi(1020)\,f_0(980)$ data for two different choices of the $f_0(980)$ properties according to Ref.~\cite{alba} (top) and \cite{nd} (bottom). 
\label{f0proper}}
\end{center}
\end{table}  

\begin{figure}[h!]
\begin{center}
\psfrag{T2}[Bb]{{\small $|T_{\phi a_0}|^2$}}
\psfrag{sqs}[tc]{{\small $\sqrt{s}$~[GeV]}}
\centerline{\epsfig{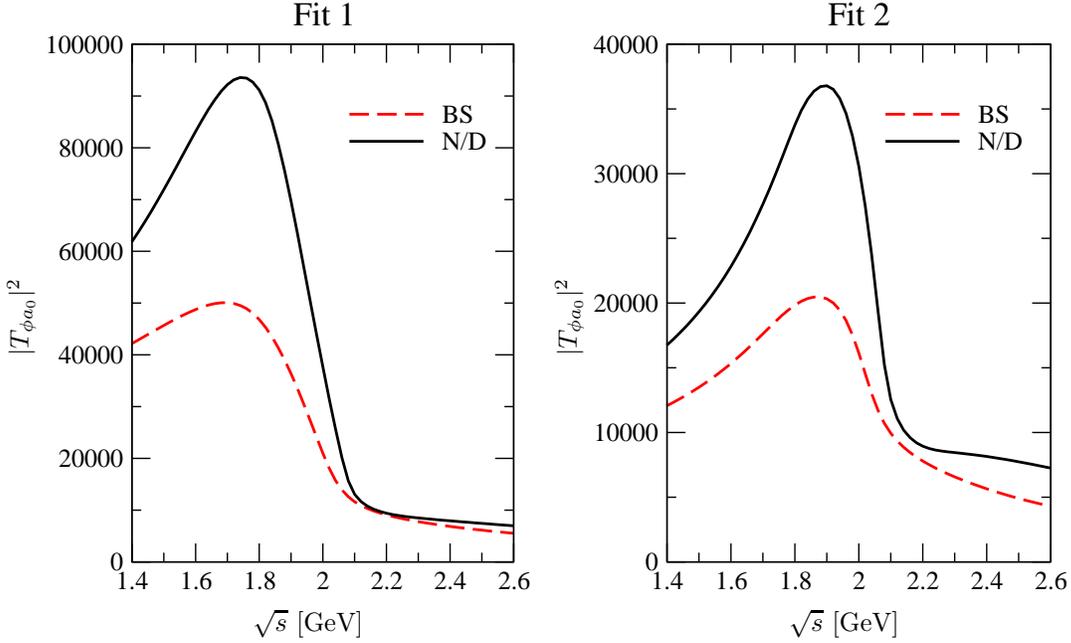}}
\vspace{0.2cm}
\caption[pilf]{\protect \small (Color online). $|T_{\phi a_0}|^2$ without local term in the kernel ${\mathcal K}_{\phi a_0}$ as a function of the $\phi a_0$ invariant mass for the possible combinations of parameters in Tables~\ref{a0proper}, \ref{f0proper}.
\label{T2_nonl}}
\end{center}
\end{figure} 
We calculate $|T_{\phi a_0}|^2$ for the four possible combinations of the parameter sets in Tables ~\ref{a0proper} and \ref{f0proper}.   As mentioned above, some of the discarded contributions to the triangle loop could modify the local term in Eq.~\eqref{vff}. For this reason, we first exclude the local contribution and concentrate on the more robust triangular topology. The $|T_{\phi a_0}|^2$ dependence on the $\phi(1020) a_0(980)$ invariant mass is shown in Fig.~\ref{T2_nonl}.  
 All the curves show a prominent enhancement below the $\phi K \bar{K}$ threshold (2.01~GeV) that hints at the presence of a dynamically generated resonance located quite close but above the $\phi \pi^0 \eta$ 
threshold (1.7~GeV). For Fit~2, the peak is narrower and has a maximum at a higher $\sqrt{s}$ but it is 2.5 times weaker than for Fit~1 (notice the different scales in the plots).

\begin{figure}[h!]
\begin{center}
\psfrag{T2}[Bb]{{\small $|T_{\phi a_0}|^2$}}
\psfrag{sqs}[tc]{{\small $\sqrt{s}$~[GeV]}}
\centerline{\epsfig{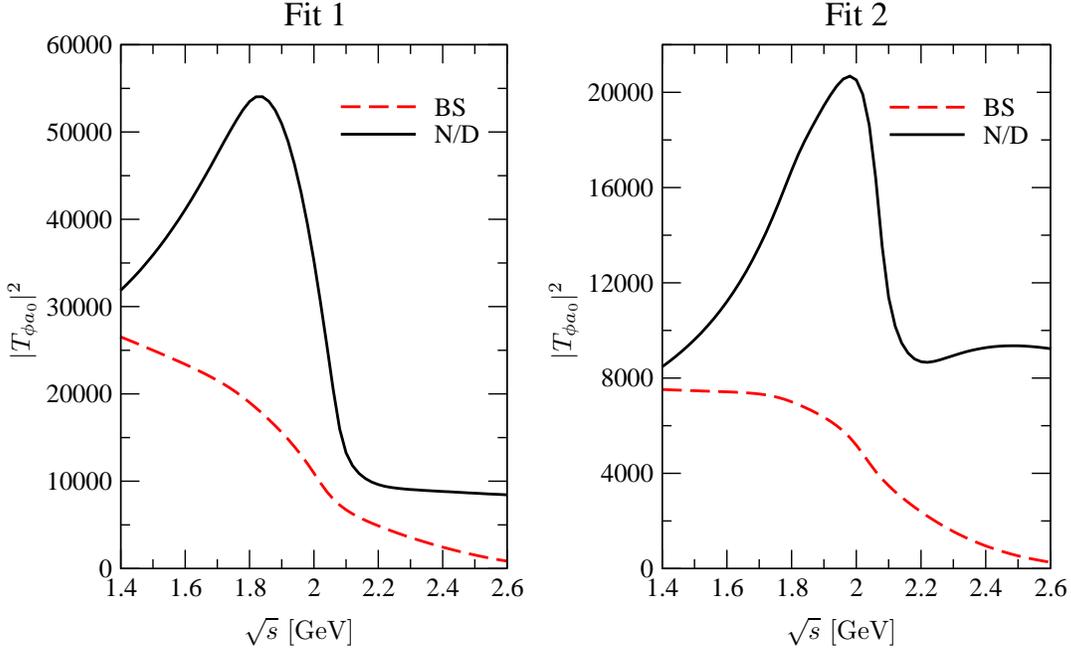}}
\vspace{0.2cm}
\caption[pilf]{\protect \small (Color online). Same as Fig.~\ref{T2_nonl} but with the local term in the kernel as in Eq.~\eqref{vff}. \label{T2}}
\end{center}
\end{figure}
Let us now take into account the local term in the kernel as given in Eq.~\eqref{vff}. For the sake of consistency the $K\bar{K}$ unitarity scalar loop function, $G_1(k^2)$, is evaluated making use of the same regularization procedure employed in generating the $a_0(980)$ resonance from Refs.~\cite{npa,nd}. 
Hence,
 when the BS set is used,  $G_1(M_{a_0}^2)$ is computed using a cut off regularization with $q_{max}=1$~GeV~\cite{npa} while,  when the N/D parameters are considered, $G_1(M_{a_0}^2)$   is obtained from a dispersion relation   with the renormalization scale fixed at the $\rho$ mass, $\mu_{K \bar{K}} = 0.77$~GeV, and a subtraction constant of $a_{KK} =-0.81$~\cite{nd}. The new results are shown in Fig~\ref{T2}. 
 In the BS case, for both Fits~1,~2, the enhancements observed before in Fig.~\ref{T2_nonl} are flatten away by the presence of the local term. This agrees with the results of Ref.~\cite{torres},  
that also makes use of the meson-meson amplitudes from Ref.~\cite{npa}, where no isovector $1^{--}$
resonance was generated. Remarkably, when the N/D set is employed the resonance peak is still clearly seen, and at a higher invariant mass with respect to Fig.~\ref{T2_nonl}, but with a $|T_{\phi a_0}|^2$  smaller by almost a factor two. Considerable differences between BS and N/D results are also observed above $\sqrt{s} > 2.2$~GeV: while the BS curve goes fast to zero, the N/D one remains nearly flat at least up to  $\sqrt{s} = 2.6$~GeV. The main difference between the two choices has to do with the actual 
value of the coupling squared $\gamma_{K\bar{K}}^2$, particularly for its imaginary part. In this way, if the BS~\cite{npa} $a_0(980)$ pole position in Table~\ref{a0proper} were used with the couplings 
of the N/D \cite{nd} pole one would obtain also $\phi(1020)a_0(980)$ broad peaks similar to those 
shown by the dashed lines in Fig.~\ref{T2_nonl}. In Ref.~\cite{fif0} it was found that the fits to BABAR~\cite{babar2} and Belle~\cite{belle} data in the region of the $Y(2175)$ resonance were stable against variation of the contact term in the $\phi(1020) f_0(980)$ kernel. Now there is more sensitivity because the $a_0(980)$ pole positions (Table~\ref{a0proper}) are not so close to the $K\bar{K}$ threshold as 
the $f_0(980)$ ones (Table~\ref{f0proper}). For this reason, the three point function $L_S$, Eq.~\eqref{ms}, is smaller than in the $f_0(980)$ case so that interferences with smaller contributions are more relevant. For the N/D \cite{nd} $a_0(980)$ pole position the local term amounts at around a 20\% of the 
leading $L_S$ contribution. However, for the BS \cite{npa} pole the corrections from the local term 
 increase significantly with energy above 2~GeV. One should notice that $G_1(M_{a_0})^2$ in Eq.~\eqref{vff} is larger by around a factor 4 for the BS pole than for the N/D one. Due to the uncertainties in the pole position and couplings of the $a_0(980)$ resonance as well as the local term 
 in ${\cal K}_{\phi a_0}$, Eq.~\eqref{vff},  we cannot arrive to a definite conclusion on the existence of an isovector companion to the $Y(2175)$ in the $\phi(1020)a_0(980)$ system. Nevertheless, we can state that if the $a_0(980)$ properties are close to those predicted by the N/D study of Ref.~\cite{nd} the present model predicts a resonance behavior of dynamical origin in the $\phi(1020) a_0(980)$ scattering around 1.8-2~GeV.\footnote{It is important to remark that the presence (or absence) of a resonance in the threshold region 
for the $a_0(980)\phi(1020)$ S-wave amplitude does not depend on the precise value of the subtraction constant $a_1$ as far as it has a natural value $-3 \lesssim a_1 < 0$. }

\begin{figure}[h!]
\begin{center}
\psfrag{K}[Bb]{{\small ${\mathcal K}_{\phi a_0}$}}
\psfrag{sqs}[tc]{{\small $\sqrt{s}$~[GeV]}}
\centerline{\epsfig{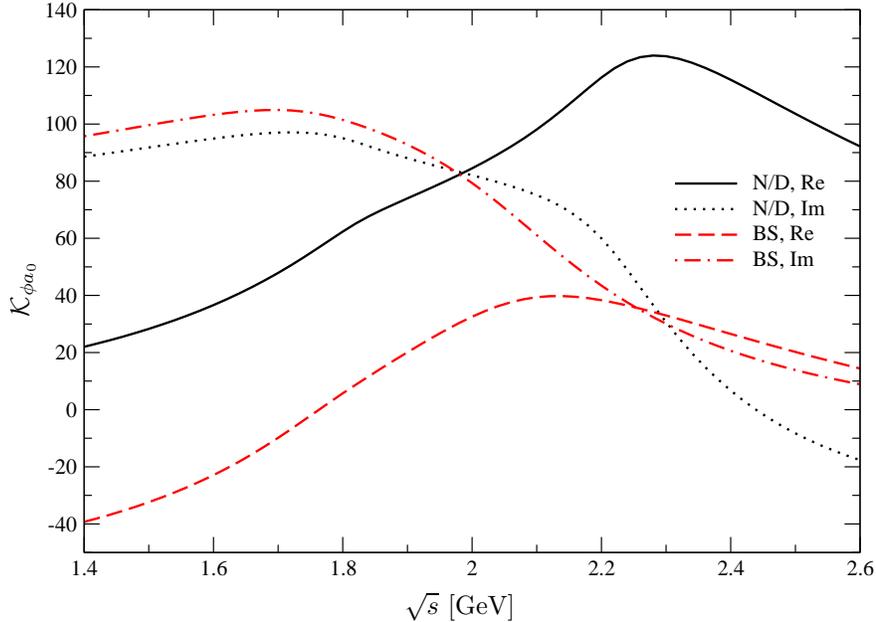}}
\vspace{0.2cm}
\caption[pilf]{\protect \small (Color online). ${\mathcal K}_{\phi a_0}$ for Fit~2 and both BS and N/D sets . \label{K}}
\end{center}
\end{figure} 
In Fig.~\ref{K} we show real and imaginary parts of the interaction potential ${\mathcal K}_{\phi a_0}$ for Fit~2 and both BS and N/D sets. In the region of $\sqrt{s}=1.6$-$2.2$, where $|T_{\phi a_0}|^2$ has a peak in the N/D case, the imaginary parts corresponding to  BS and N/D are quite similar. Instead, the real part for the N/D choice is positive (attractive) in the hole energy range of interest and larger than the BS real part, which even turns negative (repulsive) at $\sqrt{s} < 1.77$~GeV. This explains the large differences observed in $|T_{\phi a_0}|^2$. One should stress that  ${\mathcal K}_{\phi a_0}$ has an imaginary part due to a number of reasons: the finite $a_0(980)$ width, responsible for the imaginary part of the $a_0(980)$ pole position, the fact that $\gamma_{K\bar{K}}^2$ is complex, and also the imaginary part of $G_1(M^2_{a_0})$. Actually, ${\cal K}_{\phi a_0}$ should be interpreted as an optical potential.\footnote{To ensure a continuous limit to zero $a_0(980)$ width, one has to evaluate ${\cal K}_{\phi a_0}$ at the $a_0(980)$ pole position with positive imaginary part so that $k^2 \to \Rea[M_{a_0}]^2 + i \epsilon$, in agreement with Eq.~\eqref{c}. Instead, in $G_{\phi a_0}$, $M_{a_0}$ should appear with a negative imaginary part to guarantee that, in the zero-width limit, the sign of the imaginary part is the same dictated by the $i \epsilon$ prescription of Eq.~\eqref{g.cut}. Such analytical extrapolations in the masses of external particles are discussed in Refs.~\cite{barton,bar1,bar2}.}

\begin{figure}[h!]
\begin{center}
\psfrag{T2}[Bb]{{\small $|T_{\phi a_0}|^2$}}
\psfrag{sqs}[tc]{{\small $\sqrt{s}$~[GeV]}}
\centerline{\epsfig{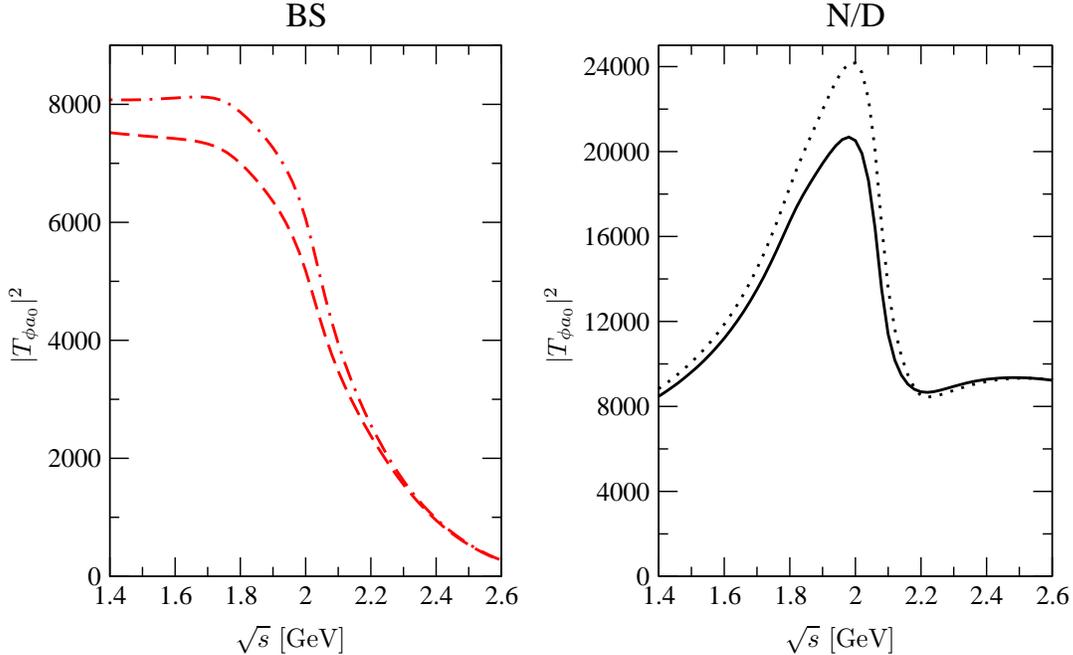}}
\vspace{0.2cm}
\caption[pilf]{\protect \small (Color online). $|T_{\phi a_0}|^2$ as a function of the $\phi a_0$ invariant mass evaluated at a fixed $a_0(980)$ pole position (dashed and solid lines) or with a convolution according to Eq.~\eqref{conv} (dash-dotted and dotted lines). All curves were obtained with Fit~2. The left panel corresponds to the $a_0(980)$ pole position of Ref.~\cite{npa} and the right one to that of Ref.~\cite{nd} (see Table~\ref{a0proper}).
 \label{T2conv}}
\end{center}
\end{figure}

So far, the $a_0(980)$ pole position has been used as a complex value for the  $M_{a_0}$ mass. It is instructive to calculate the amplitude squared taking  instead a convolution over the $a_0(980)$ mass distribution determined by its width, so that only real masses appear now in $G_{\phi a_0}$, which has then its cut along the real axis above threshold, as required by two-body unitarity with real masses. Namely, we calculate
\begin{equation}
|T_{\phi a_0}|^2_{\mathrm{conv}}(s) =\frac{1}{N} \int^{\Rea(M_{a_0}) + \delta}_{\Rea(M_{a_0}) - \delta} dW   \frac{\Ima(M_{a_0})}{[W-\Rea(M_{a_0})]^2 + \Ima(M_{a_0})^2} |T_{\phi a_0}(s,M_{a_0},W)|^2 \,, 
\label{conv}
\end{equation}
with $T_{\phi a_0}(s,M_{a_0},W)$ defined as
\begin{equation}
T_{\phi a_0}(s,M_{a_0},W)= \frac{{\cal K}_{\phi a_0}(s,M_{a_0})}{1+{\cal K}_{\phi a_0}(s,M_{a_0}) G_{a_0 f_0}(s,W)}\,,
\end{equation}
and the normalization
\begin{equation}
N=\int^{\Rea(M_{a_0}) + \delta}_{\Rea(M_{a_0}) - \delta} dW   \frac{\Ima(M_{a_0})}{[W-\Rea(M_{a_0})]^2 + \Ima(M_{a_0})^2} \,.
\end{equation}
 $\Rea(M_{a0})$ and $\Ima(M_{a_0})$ are the real and (positive) imaginary part of the $a_0(980)$ pole position.
The integration interval around the maximum of the distribution, characterized by $\delta$, should be enough to cover the region where the $a_0(980)$ strength is concentrated. In Fig.~\ref{T2conv} we compare the results obtained in this way with those obtained from Eq.~\eqref{tff} at a fixed complex $M_{a_0}$. This is done for Fit~2, both BS and N/D parameters and using $\delta = 5 \,\Ima(M_{a_0})$. Only small differences arise in the hight of the peak so that one can conclude that the two approaches produce the same qualitative features, as one would expect based on physical reasons.  

\subsection{$\phi(1020) \, a_0(980)$ scattering corrections to $e^+e^-\to\phi(1020)\pi^0\eta$}
\label{sec:fsi}

\begin{figure}[h]
\begin{center}
\psfrag{sigma (nb)}[Bb]{{\small $\sigma$~[nb]}}
\psfrag{sqs}[tc]{{\small $\sqrt{s}$~[GeV]}}
\centerline{\epsfig{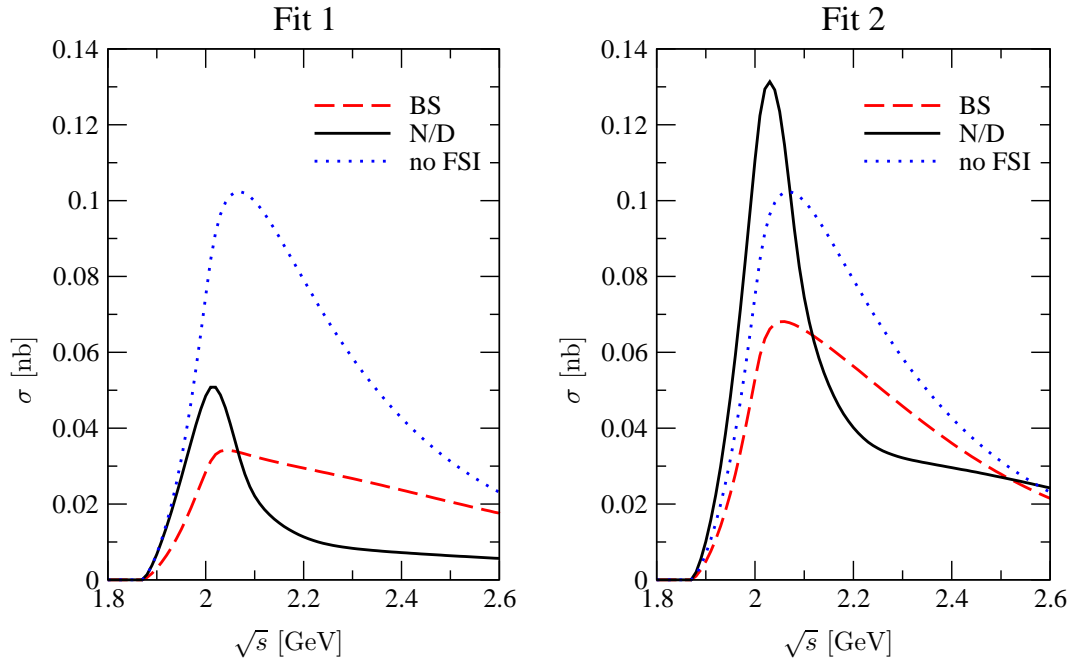}}
\vspace{0.2cm}
\caption[pilf]{\protect \small (Color online). $e^+e^-\to \phi(1020)\pi^0\eta$ cross section in the $a_0(980)$ region 
($M_{\pi\eta}\in[0.85,1.10]$~GeV). The dotted line in both plots is the result of Ref.~\cite{araujo} where final state $\phi(1020) a_0(980)$ re-scattering was not considered. The rest of the lines include  FSI according to Eq.~\eqref{fsi}
for the sets of parameters given in Tables~\ref{a0proper},~\ref{f0proper}. 
\label{fig:cross}}
\end{center}
\end{figure}

The findings described above have direct implications for the $e^+e^-\to \phi(1020)\pi^0\eta$
reaction with the $\pi^0\eta$ invariant mass in the $a_0(980)$ mass region.\footnote{Here, for simplicity, we identify the $\eta_8$ state with the physical $\eta$ particle, neglecting $\eta_8-\eta_1$ mixing. This is also done in Refs.~\cite{npa,nd}, from where the meson-meson scattering amplitudes in the $a_0(980)$ channel have been obtained. New studies indicate that the $a_0(980)$ coupling
to $\pi\eta'$ is very small~\cite{upcoming}.} This process has been investigated in Ref.~\cite{araujo} where the presence of the $a_0(980)$ is properly taken into account by replacing the lowest order $K \bar{K} \to \pi \eta$ tree level vertex 
from ${\cal L}_2$ Eq.~\eqref{lag2} by the unitarized amplitude of Ref.~\cite{npa}. However, the corrections due to $\phi(1020)a_0(980)$ re-scattering (FSI) were not included. Here we consider the impact of these FSI on the total $e^+e^-\to \phi(1020)\pi^0\eta$ cross section using the previously derived $\phi(1020)a_0(980)$ amplitude. Under the assumption that the $e^+e^-\to \phi(1020)\pi^0\eta$ reaction is dominated by the  $\phi(1020)a_0(980)$ channel, the cross section after FSI can be cast as~\cite{fif0,ddecays,gamma}
\begin{align}
\sigma_{FSI}= \sigma_0 \Biggl|\frac{1}{1+{\cal K}_{\phi a_0}(s)G_{\phi a_0}(s)}\Biggr|^2\,.
\label{fsi}
\end{align} 
We take $\sigma_0$ from Ref.~\cite{araujo} (Fig.~5), which was obtained by integrating the $\pi \eta$ invariant mass $M_{\pi \eta}$ in the $a_0(980)$ region (850-1100~MeV) so that our assumption of $\phi a_0$ dominance is justified. 
The results are shown in Fig.~\ref{fig:cross} for the different parameter sets.     
We find considerable FSI corrections. In particular, for Fit~1 the reduction of the cross section is large, even a factor five at some energies. With the BS choice, the cross section does not exhibit any structure and is smoother than the one without FSI. Instead, for the N/D set a peak (quite prominent for Fit~2) is observed at $\sqrt{s}\sim2.03$~GeV. These results clearly show the 
interest of measuring experimentally the $\pi\eta$  invariant mass distribution so as to confirm the existence of this new isovector $J^{PC}=1^{--}$ resonance that would be observed as a clear peak in data. The existence of this resonance is favored by our results since it appears when the $a_0(980)$ properties from the later and more complete N/D~\cite{nd} calculation are adopted.

%%%%%%%%%%%%%%%%%%%%%%%%%%%%%%%%%%%%%%%%%%%%%%%%%%%%%%%%%%%%%%%%%%%%%%%%%
%%%%%%%%%%%%%%%%%%%%%%%%%%%%%%%%%%%%%%%%%%%%%%%%%%%%%%%%%%%%%%%%%%%%%%%%%
\section{Summary and conclusions}
\label{sec:conclu}

We have studied the $I=1$ S-wave $\phi(1020) a_0(980)$ dynamics around threshold paying special attention to the possible  
dynamical generation of an isovector $J^{PC}=1^{--}$ scalar resonance. Following the approach of Ref.~\cite{fif0}, where 
the related isoscalar S-wave $\phi(1020) f_0(980)$ scattering was investigated, we first considered the scattering of the $\phi(1020)$ resonance with a pair of light pseudoscalar mesons at tree level using chiral Lagrangians coupled to vector mesons by minimal coupling.  The re-scattering of the two pseudoscalars in $I=1$ and S-wave  generates dynamically the $a_0(980)$. We have used the information about this state (pole position and residue in the $K \bar{K}$ channel) from two different studies of meson-meson scattering in coupled channels to determine the $\phi(1020) a_0(980)$ scattering potential without introducing new extra free parameters. Afterwards the full amplitude is obtained by resummation of the $\phi(1020) a_0(980)$ unitarity loops. The parameter $g^2$, characterizing $\phi(1020) K$ scattering at threshold, and the $\phi a_0$ subtraction constant $a_1$ are obtained from two different fits to $e^+e^-\to\phi(1020) f_0(980)$ BABAR~\cite{babar2} and Belle~\cite{belle} data. We find that if the physical $a_0(1980)$ properties correspond to those extracted with the N/D method in Ref.~\cite{nd} (see Table~\ref{a0proper}), the present model predicts a resonance of dynamical origin around 1.8-2~GeV.  A broader resonance is also generated when the $a_0(980)$ pole position and couplings are taken from the BS study of Ref.~\cite{npa} if the strength of the local term in the $\phi(1020) a_0(980)$ interaction kernel is reduced.

Furthermore, we have determined the $\phi(1020) a_0(980)$ final state interactions that strongly modify 
the cross section for the reaction $e^+e^-\to\phi(1020)\pi^0\eta$ when the $\pi^0\eta$ invariant mass is in the $a_0(980)$ region.  If the $a_0(980)$ properties from the N/D method are taken, a strong clearly   visible peak around 2.03~GeV is observed, signaling the presence of the dynamically generated isovector $1^{--}$ resonance. For the $a_0(980)$ BS pole of Ref.~\cite{npa} no peak is generated but a strong reduction of the cross-section takes place.  The present results further support the idea that a study of the $e^+e^- \to \phi(1020) a_0(980)$ reaction, which should be accessible at present $e^+e^-$ factories~\cite{babar1,bes08,belle}, may provide novel relevant information about hadronic structure and interactions in the 2~GeV region.

%%%%%%%%%%%%%%%%%%%%%%%%%%%%%%%%%%%%%%%%%%%%%%%%%%
\section*{Acknowledgements}
We thank Mauro Napsuciale and Carlos Vaquera-Araujo for sending us their results corresponding to the dotted lines in Fig.~\ref{fig:cross}, and Eulogio Oset for useful discussions. 
This work has been partially funded by the MEC grant FPA2007-6277 and Fundaci\'on S\'eneca grant 11871/PI/09.  We also acknowledge the financial support from  the BMBF grant 06BN411, EU-Research Infrastructure Integrating Activity "Study of Strongly Interacting Matter" (HadronPhysics2, grant n. 227431)
under the Seventh Framework Program of EU and   the Consolider-Ingenio 2010 Program CPAN (CSD2007-00042).

\end{document}